# A Hybrid Deep Learning Classification of Perimetric Glaucoma Using Peripapillary Nerve Fiber Layer Reflectance and Other OCT Parameters from Three Anatomy Regions

Keywords:

Authors: Ou Tan,[1] David S. Greenfield[3]; Brian A. Francis[4]; Rohit Varma[5]; Joel S. Schuman[6]; David Huang[1], Dongseok Choi[1,2]

**Institute:** [1]Casey Eye Institute, Oregon Health & Science University; [2]OHSU-PSU School of Public Health, [3]Bascom Palmer Eye Institute, University of Miami; [4]Doheny Eye Center and David Geffen School of Medicine at UCLA; [5]Southern California Eye Institute; [6] Wills Eye Hospital

Corresponding Author: Dongseok Choi, PhD

Funding Sources: NIH grants R21 EY032146, R01 EY023285, and P30 EY010572, and an unrestricted grant from Research to Prevent Blindness to Casey Eye Institute.

Commercial relationships: OT: P; DH: F,R,I,P; DC: None; Others: None

# Abstract

**Précis:**

A hybrid deep-learning model combines NFL reflectance and other OCT parameters to improve glaucoma diagnosis.

**Objective:**

To investigate if a deep learning model could be used combine nerve fiber layer (NFL) reflectance and other OCT parameters for glaucoma diagnosis.

**Patients and Methods:**

This is a prospective observational study where of 106 normal subjects and 164 perimetric glaucoma (PG) patients. Peripapillary NFL reflectance map, NFL thickness map, optic head analysis of disc, and macular ganglion cell complex thickness were obtained using spectral domain OCT. A hybrid deep learning model combined a fully connected network (FCN) and a convolution neural network (CNN) to develop to combine those OCT maps and parameters to distinguish normal and PG eyes. Two deep learning models were compared based on whether the NFL reflectance map was used as part of the input or not.

**Results:**

The hybrid deep learning model with reflectance achieved 0.909 sensitivity at 99% specificity and 0.926 at 95%. The overall accuracy was 0.948 with 0.893 sensitivity and 1.000 specificity, and the AROC was 0.979, which is significantly better than the logistic regression models (p < 0.001). The second best model is the hybrid deep learning model w/o reflectance, which also had significantly higher AROC than logistic regression models (p < 0.001). Logistic regression with reflectance model had slightly higher AROC or sensitivity than the other logistic regression model without reflectance (p = 0.024).

**Conclusions:**

Hybrid deep learning model significantly improved the diagnostic accuracy, without or without NFL reflectance. Hybrid deep learning model, combining reflectance/NFL thickness/GCC thickness/ONH parameter, may be a practical model for glaucoma screen purposes.

# 1 Introduction

Glaucoma is a leading cause of blindness, and about half of the patients do not know that they have it. Peripapillary nerve fiber layer (NFL) thickness measurement by optical coherence tomography (OCT) has been widely used in the clinical management of glaucoma.[1-8] Overall NFL thickness average is useful for confirming the diagnosis of glaucoma. But its diagnostic sensitivity is not sufficient to be used alone for population-based screening.[9,10] At the 99% specificity, the best NFL thickness parameter has a sensitivity of only 20-60% for perimetric glaucoma (PG), which is too low for diagnostic screening.[11-16]

In early glaucoma, animal studies showed that NFL thinning lagged behind axonal loss on a time scale of months.[17] NFL reflectivity was reduced in glaucoma subjects,[18] presumably due to loss of axons and axonal microtubule content.[19-21] It was expected to find loss in NFL reflectivity defect earlier than NFL thickness. Though the average NFL reflectivity, as a diagnostic parameter, underperformed the average NFL thickness,[18] the diagnostic accuracy was improved by normalizing the NFL reflectivity by one or more outer retinal layers.[22,23] The combination of the normalized reflectivity with NFL thickness further improves the diagnostic accuracy.[23,24] Our preliminary results showed that the focal reflectance loss had significantly higher (p=0.017) diagnostic accuracy (area under receiver operating characteristic curve, AROC= 0.925) than the overall average NFL thickness (0.859).[25]

Combining OCT parameters from different anatomy areas improved the diagnostic accuracy.[11,13,26-29] We had previously developed glaucoma structural diagnostic index (GSDI), a combination of OCT parameters from 3 anatomic regions: disc, peripapillary retina, and macula. The parameters were combined by logistic regression. We showed that the GSDI produced higher diagnostic accuracy than any parameter from a single region.[11] However, 31% of the perimetric glaucoma eyes were still missed at the 99% specificity level.

In this paper, we further improved the classification accuracy of PG against normal control subjects by combining nerve fiber layer (NFL) reflectance map and thickness map in a convolution neural network and a fully connected network of clinical and ocular parameters in a hybrid deep learning model.

# 2 Methods

## 2.1 Participants

In this secondary data analysis (NIH R21 EY032146) of the Advanced Imaging for Glaucoma (AIG) study, 620 scanning data from 106 normal subjects and 671 scanning data from 164 perimetric glaucoma patients were used to develop a hybrid deep learning model to classify PG patients from normal subjects. All data were from the baseline visits from the AIG study. The details of the AIG study were published previously.{Le, 2015 #65} The study followed the Declaration of Health Insurance Portability and Accountability Act of 1996 (HIPAA) privacy and security regulations. All participating Institutional review boards approved the study, and written informed consent was obtained from all patients. This study was approved by the Institutional Review Board (IRB) of Oregon Health & Science University.

Eyes enrolled in the PG group had glaucomatous optic neuropathy as evidenced by diffuse or localized thinning of the neuroretinal rim or NFL defect on fundus examination, and corresponding repeatable VF defects with PSD (P < .05) or GHT outside normal limits. Both eyes of normal participants met the following criteria: VF tests within normal limits, IOP<21 mm Hg, and normal optic nerve on slit-lamp biomicroscopy. Exclusion criteria common to all groups included best-corrected visual acuity (BCVA) worse than 20/40, evidence of retinal pathology, or history of keratorefractive surgery.

## 2.2 Data acquisition

Participants were scanned with a spectral-domain OCT systems (RTVue, Visionix/Optovue, Inc., Fremont, CA, USA) from three clinic centers of the AIG study. Both eyes of each participant were scanned with the ONH scan and GCC scan twice or three times. The ONH scan was a 4.9 mm composite scan that centers on the disc and contains 13 concentric scans covering the peripapillary region, and 12 radial scans covering the disc region. GCC scan is 7mm raster scan covering the macula. RTvue software provided 1) RNFL thickness analysis(NFL thickness profile at D=3.4mm), the NFL thickness map, and ONH analysis for ONH scans; 2) macular GCC thickness for GCC scans. {Loewen, 2015 #68}

## 2.3 Nerve Fiber layer reflectance

We adapted the NFL reflectance map calculation from the cubic scan to the ONH scan.[25] In short, the reflectance was summed in the NFL band, from 7 pixels under ILM to the outer NFL boundary. Then, the summation was normalized by the reflectance averaged in the photoreceptor and pigment epithelium complex (PPEC) band. The reflectance values shadowed by the vessel were replaced with neighboring pixels to preserve continuity. The difference from previous flowchart is that the NFL reflectance map was interpolated from NFL reflectance ratio profiles from 13 rings.

After re-centering the map to the disc center, we selected a 2.1 – 4.2 mm diameter analytic zone. The region outside the 4.2-mm diameter was excluded to avoid cropping artifacts from possible scan de-centration, while the region inside the 2.1-mm diameter was excluded to mask the optic disc. Then, we performed the azimuthal spatial frequency filtering to remove the first-degree angular component in the azimuthal dimension, as we found that the azimuthal filter reduced the bias caused by the incident angle.[25]

The filtered NFL reflectance map was divided into superpixels. The superpixel grid contains 32 tracks parallel to the average nerve fiber trajectory map. Each track was evenly divided into three segments in the annulus between 2.1 and 4.2 mm from the center of the disc. The NFL reflectance in each superpixel was averaged.

We also obtained superpxiel values for NFL thickness map following the same method.

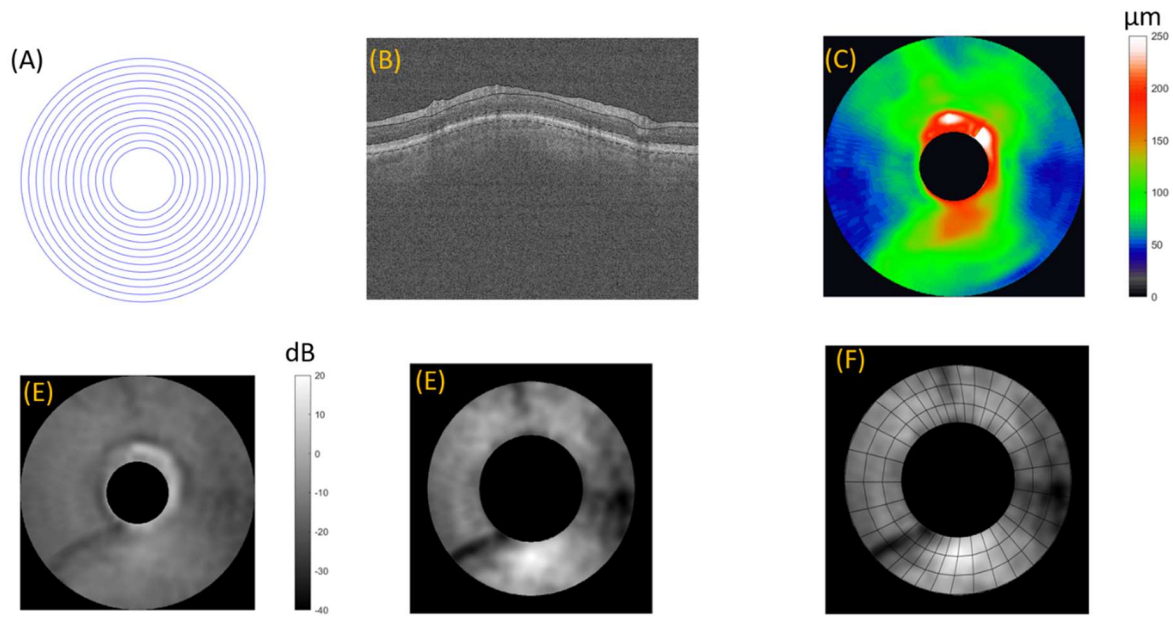

Figure 1. Nerve fiber layer (NFL) thickness and reflectance maps (A) Scan included 13 concentric rings around optio disk; (B) NFL thickness and NFL reflectance were obtained from NFL, the top bright band; NFL reflectance is also normalized by the retinal pigment epimembrance (RPE); (C) NFL thickness map were reconstructed from the NFL thickness profile of 13 rings; The eye is a glaucoma eye; (D) NFL reflectance map of same eye; (E) NFL reflectance map was filtered to reduce the variance due to incident angle; (F) the superpixel grid used on maps to reduce the dimension (from 655*655 map to 32*3 grid).

## 2.4 Hybrid deep learning models

The hybrid deep learning model combined a fully connected network (FCN) and a convolution neural network (CNN) to distinguish normal and PG eyes(Figure 2 ). A 33 X 3 grid was used for the input of CNN. Two channels were used in the CNN, one corresponding to NFL thickness, and the other corresponding to NFL reflectance. Since both the NFL reflectance and thickness maps were on a 32x2 concentric circular grid, the maps were padded similarly to Shubert et al.[30]

In the FCN, OCT parameters from disc and macula were used. The macular GCC thickness parameters were obtained from GCC scan. We only used superior and inferior hemisphere average, and GCC focal loss volume(FLV). The disc parameters were obtained from the ONH analysis, including disc area, rim area, cup-disc-area-ratio, vertical cup-disc –ratio (VCDR). In addition, age, gender, and axial length were included in the FCN.

To compare the performance with or without NFL reflectance, we also trained a similar hybrid deep learning model with one channel CNN of superpxiel values of NFL thickness only.

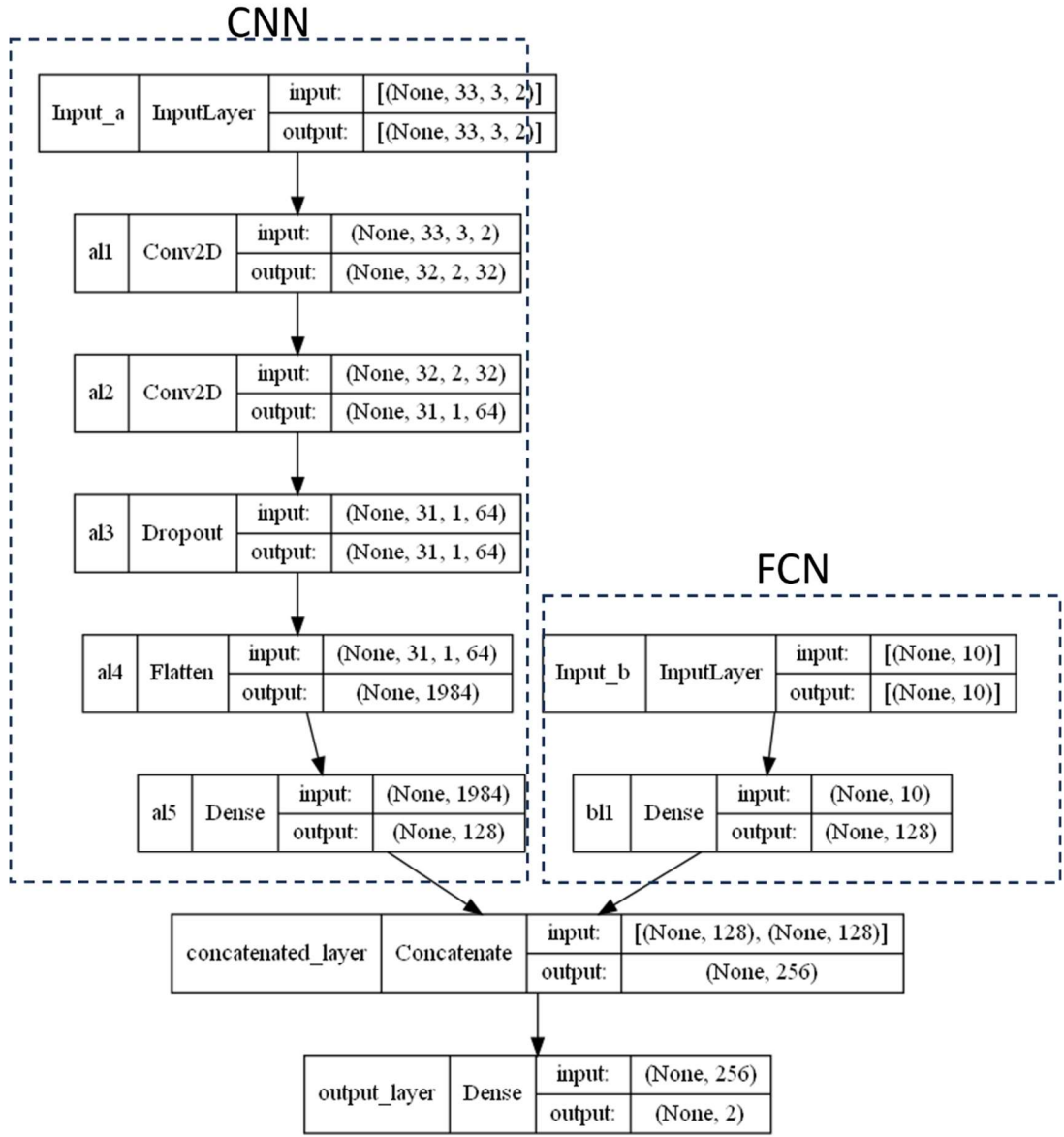

**Figure 2**. Hybrid deep learning model architecture used a convolutional neural network (CNN) to encode the NFL reflectance and thickness superpixels and a fully connected network (FCN) to encode OCT demography and OCT parameters from the macula and optic nerve head.

## 2.5   Other machine learning models

We also trained other machine learning models. The first one was a logistic regression model with RNFL thickness overall average, RNFL FLV (obtained on NFL thickness profile at 3.4mm circle), GCC thickness & FLV, and VCDR, which were same as we used in our previous publicaton.{Loewen, 2015 #68} In the second logistic regression model, two more OCT variabels, NFL relectance average and  FLV from the peripiallry area (2.1-4.2mm),were added to the first logistic regression model. Age and axial length were included as random- effects and gender as a fixed-effect in both models.

## 2.6  Training and Testing

The study subjects in each group were split into training (80%) and test sets (20%) to train the deep learning model in Figure 1. In training, the batch size of 600, 3500 epochs, 2x2 kernel size, and a validation split of 0.2 was used with the Adam optimizer, a learning rate of 0.00008, and binary cross-entropy loss.

## 2.7  Statistical analysis

All computations were done using the pROC, lmerTest, keras and tensorflow packages in the statistical language R.[31] The area under the receiver operating characteristic curve (AROC) and sensitivity at 95% and 99% specificity of the testing data were compared among the models.

# 3  Results

Two hundred and ten normal eyes from 106 participants and 238 PG eyes from 164 participants were selected for this study from the AIGS study dataset (table 1). The participants provided, on average, 4.8 scans in this cross-sectional subset of the baseline data from the AIG study. The overall average age was 62.3 (standard deviation 9.76), and 61.9% were female. There were no significant differences in age and ratio of females between the normal and PG groups (t-test, p-value = 0.152 and 0.312, respectively).

Table 1. Clinical and ocular characteristics of subjects

|  |  | Normal (n=106) | Perimetric Glaucoma (n =164) |
|---|---|---|---|
| Demography | Age (years) | 59.8 ± 9.72 | 63.8 ± 9.49 |
|  | Female | 70 (66.0%) | 97 (59.1%) |
|  | Axial length | 23.7 ± 1.03 | 24.4 ± 1.32 |
|  | Visual Field Mean Deviation | -0.10 ± 1.02 | -4.57 ± 3.96 |
| OCT parameters | Disc area | 2.11 ± 0.348 | 2.17 ± 0.450 |
|  | Rim area | 1.42 ± 0.269 | 0.904 ± 0.357 |
|  | Cup to disk ratio | 0.315 ± 0.143 | 0.569 ± 0.184 |
|  | Cup to disk vertical ratio | 0.510 ± 0.156 | 0.751 ± 0.162 |
|  | GCC SUP | 94.9 ± 6.78 | 84.2 ± 10.6 |
|  | GCC INF | 96.1 ± 7.09 | 80.4 ± 11.6 |
|  | GCC FLV | 0.735 ± 0.955 | 5.34 ± 3.91 |
|  | RNFLT AVG | 99.2 ± 8.53 | 80.1 ± 11.8 |
|  | RNFLT FLV | 1.72 ± 1.90 | 7.71 ± 4.54 |

| | | |
|---|---|---|
| NFLR AVG | -8.11 ± 1.29 | -11.6 ± 2.26 |
| NFLR FLV | -0.219 ± 0.375 | -2.86 ± 2.11 |

* GCC SUP=ganglion cell complex superior hemisphere thickness; GCC INF: GCC inferior hemisphere thickness; GCC FLV: GCC focal loss volume; RNFLT: nerve fiber layer thickness at 3.4mm circle; NFLT: nerve fiber layer thickness in ring area 2.1~4.2mm; NFLR: nerve fiber layer reflectance.; AVG: overall Average.

About 80% of the subjects in each group (968 scanning data from 216 subjects) were used for training, and the rest (235 scanning from 54 subjects) were used to evaluate the trained model performance. The trained hybrid deep learning model with reflectance achieved 0.945 overall accuracy with 0.900 sensitivity and 0.990 specificity. The area under the receiver operating characteristics curve (AROC) was 0.985. The sensitivities were 0.899 and 0.952 at 99% and 95% specificities, respectively.

With the test data (Table 2), the hybrid deep learning model with reflectance achieved 0.909 sensitivity at 99% specificity and 0.926 at 95%. The overall accuracy was 0.948 with 0.893 sensitivity and 1.000 specificity, and the AROC was 0.979, which is significantly better than the logistic regression models (p < 0.001). The second-best model is the hybrid deep learning model w/o reflectance, which also had significantly higher AROC than logistic regression models (p < 0.001). Logistic regression with reflectance model had slightly higher AROC or sensitivity than the other logistic regression model without reflectance (p = 0.024).

**Table 2. Test performance of diagnostic accuracy**

| | | Sensitivity | |
|---|---|---|---|
| | AROC | at 95% Specificity | at 99% specificity |
| Logistic regression w/o reflectance | 0.923 | 0.813 | 0.718 |
| Logistic regression with reflectance | 0.931+ | 0.852 | 0.708 |
| Hybrid deep learning model thickness alone | 0.978* | 0.917 | 0.901 |
| Hybrid deep learning model with thickness and reflectance | 0.979* | 0.926 | 0.909 |

+p = 0.024 against logistic regression w/o reflectance

*p < 0.001, against both logistic regression models

## 4  Discussion

In this study, we proposed a hybrid deep learning model to combine peripapillary NFL reflectance and NFL thickness, macular GCC thickness, and OCT parameters for ONH for glaucoma diagnosis. The multi-modal hybrid deep learning model significantly improved the diagnostic accuracy compared to simple logistic regression models and reached a level suitable for glaucoma screen purposes.

Previously, we showed that combining OCT parameters from 3 anatomy regions using logistic regression had significantly higher diagnostic accuracy (AROC=0.92) than the best single OCT parameter (AROC=0.90).[11] Using the same population, the hybrid deep learning models had significantly better AROC (AROC=0.98) or sensitivity(0.90~0.91) at 99% specificity than the simple logistic regression model(AROC=0.92, sensitivity=0.71~0.72), with/without reflectance. The improved performance is because 1) the deep learning model was more efficient in learning the loss pattern of NFL thickness or reflectance in the 32x3 superpixel grid than the logistic regression model, in which the focal loss was simply integrated in a significant loss mask; 2) the confounding factors, were used for nonlinear adjustments in the deep learning model, might further reduce the bias due to individual variance.

Other researches also reported that combining multiple OCT parameters using machine learning significantly improved the diagnostic accuracy, compared to single best OCT parameter.[32] Some reported even higher AROC (AROC>0.985) or sensitivity than this study using random forest[33], using logistic regression on 16 OCT parameters,[29] or using regression tree.[34] But either the average glaucoma stages in these studies were more severe than this study VF-MD=-5.6 dB[33] and –9.0 dB[34]), or the best single parameter had much higher accuracy itself(AROC =0.987[29]). So the AROC difference between machine learning and the best single parameter in those models (0.02-0.04 increase of AROC) was actually less than in this study (~0.08 increase of AROC). Also several studies showed than using deep learning models may improve the diagnostic accuracy for glaucoma using OCT.[35-38] Due to the differences in population, OCT device, and selected OCT parameters among studies, it is not simple to compare the performance among models just based on these AROC differences. However, it showed that the hybrid deep learning model is an efficient way to combine OCT parameters to improve glaucoma diagnostic accuracy.

Other deep learning models were developed to improve glaucoma diagnosis using NFL reflectance.[39] in this study, en face projection, which is equivalent to average NFL reflectivity without normalization. Though the deep learning model significantly improved the diagnostic accuracy than traditional OCT, adding the en face projection actually reduced the performance of the deep learning model.

In this study, Adding NFL reflectance to the combination of multiple OCT parameters only slightly improved the diagnostic accuracy for both logistic regression and deep learning methods. In another study, we showed that NFL reflectance may significantly improve diagnostic accuracy.{Tan, 2021 #120} Leung also reported better performance of NFL optical texture analysis than NFL thickness., in which the NFL optical texture was similar to the normalized NFL reflectance in this study, but with more optimization.[40] One main difference between this study with other studies is the data quality. This study was based on an OCT with a 27kHz scanning rate and a scan with 13 circles to cover the 4.9mm parapapillary NFL area, in which the eye motion between circular scans was not corrected. Our another study was based on the OCT with a 100kHZ scanning rate and a 4.5mm cubic scan with the eye motion corrected. Leung's study is based on a 100kHz swept source OCT and a 12*9mm cubic scan.[41] A denser scan and less eye motion led to less variance of incident angle and better performance of azimuthal spatial frequency filtering, which benefits NFL reflectance more than NFL thickness because NFL reflectance is highly sensitive to incident angle.

# 5 Conclusion

In summary, adding NFL reflectance to combination of NFL thickness, GCC thickness and cup disc ratio helped to improve the diagnostic accuracy, but not significantly. Hybrid deep learning model significantly improved the diagnostic accuracy, without or without NFL reflectance. Hybrid deep learning model, combining reflectance/NFL thickness/GCC thickness/ONH parameter, may be a practical model for glaucoma screen purposes.

# Acknowledgements

This study was supported by NIH grants R21 EY032146, R01 EY013516, R01 EY023285, P30 EY010572, the Malcolm M Marquis, MD Endowed Fund for Innovation and an unrestricted grant from Research to Prevent Blindness to Casey Eye Institute. The sponsor or funding organization had no role in the design or conduct of this research.